\documentclass[11pt]{amsart}
\usepackage[a5paper]{geometry}

\usepackage{mathrsfs}
\usepackage{amsfonts}
\usepackage{amsmath} 
\usepackage{amssymb} 
\usepackage{amsthm}
\usepackage{graphicx} 
\usepackage{url}
\usepackage{color}
\usepackage{hyperref}
\definecolor{refcolor}{RGB}{0,0,190}
\hypersetup{
    colorlinks,
    citecolor=refcolor,
    filecolor=refcolor,
    linkcolor=refcolor,
    urlcolor=refcolor
}

\theoremstyle{definition}

\def\({\left(}
\def\){\right)}

\newcommand{\tn}{\textnormal}

\newcommand{\hilbert}{\mathcal{H}}

\newcommand{\mc}[1]{\mathcal{#1}}

\newcommand{\Prj}{\mathcal{P}}
\newcommand{\abs}[1]{\left|#1\right|}

\newcommand{\ie}{\textit{i.e.} }
\newcommand{\vs}{\textit{vs.} }

\newcommand{\etc}{\textit{etc}}

\newcommand{\U}{\tn{U}}

\newcommand{\schrod}{Schr\"odinger}

\newcommand{\ks}{Kochen-Specker}
\newcommand{\leggarg}{Leggett-Garg}
\newcommand{\bra}[1]{\langle#1|}
\newcommand{\ket}[1]{|#1\rangle}

\newcommand{\expectation}[1]{\langle#1\rangle}

\def\sref #1{\S\ref{#1}}

\newcommand{\Law}[1]{#1}
\newcommand{\PProj}{\Law{PROJ}}
\newcommand{\PProb}{\Law{PROB}}

\newcommand{\TENS}{\Law{TENS}}

\title{A principle of quantumness}

\author{Ovidiu Cristinel Stoica*}
\thanks{*Department of Theoretical Physics, National Institute of Physics and Nuclear Engineering -- Horia Hulubei, Bucharest, Romania. Email: \href{mailto:cristi.stoica@theory.nipne.ro}{cristi.stoica@theory.nipne.ro}}
\date{\today}

\begin{document}

\begin{abstract}
Quantum correlations and other phenomena characteristic to a quantum world can be understood as simply consequences of a principle derived from the postulates of Quantum Mechanics. This explanatory principle states that these phenomena specific to the quantum world are caused by the tension between the constraints, or initial conditions, imposed by incompatible observations. This tension is found to be at the root of Bohr's complementarity, Heisenberg's uncertainty, results concerning nonlocality, contextuality, quantum correlations in time and space. This tension requires the presence of noncommuting observables, but noncommutativity doesn't always lead to the tension, and the two concepts are not exactly the same, as it will be explained and exemplified.
\end{abstract}

\maketitle

\section{Introduction}

In the following, I will call ``quantumness'' those distinctive characteristics of Quantum Mechanics (QM) which make it different from classical physics. Although these phenomena are predicted by the theory itself, they still appear strange and contradictory, and there is a spread opinion that they lack an explanation. While the mere fact that Quantum Mechanics {\em predicts} them, means that it {\em explains} them in terms of its fundamental postulates, this doesn't make them any clearer. This article aims to identify the key feature that is at the root of all these phenomena. By this, what appears as the strangeness of QM can at least be reduced to a single strange feature.

Many physicists consider desirable to be able to isolate the source of quantumness in the form of a principle, more like how Special Relativity is founded on two simple principles that have a physical meaning. I emphasize that this doesn't necessarily mean to search for a principle outside what QM already tells us, but only to isolate the essence, the root of all phenomena that make QM so different from classical physics.

With respect to this, J.A. Wheeler wrote \cite{Wheeler1979QuantumAndUniverse}
\begin{quotation}
... if one really understood the central point and its necessity in the construction of the world, one ought to be able to state it in one clear, simple sentence. Until we see the quantum principle with this simplicity we can well believe that we do not know the first thing about the universe, about ourselves, and about our place in the universe.
\end{quotation}
According to Fuchs and Stacey \cite{Fuchs2014NegativeRemarks},
\begin{quotation}
Can we find some axiomatic system that really goes after the {\em weird} part of quantum theory? [...]
What I would like as a goal is a way to push quantum theory's specific form of contextuality all into one corner.
\end{quotation}

The aim of this article is to make more explicit the fact that quantumness is simply a consequence of a principle which I will call {\em the tension principle}, and which follows directly from the projection postulate. Consequently,
\begin{quotation}
``In one clear, simple sentence''\cite{Wheeler1979QuantumAndUniverse}, the ``{\em weird} part of quantum theory''\cite{Fuchs2014NegativeRemarks} is the tension between the constraints imposed to the system by different observations.
\end{quotation}

From the beginning of Quantum Mechanics, starting with Bohr and Heisenberg, physicists were aware of the fact that noncommuting observables are incompatible, or complementary. However, while the tension principle is related to noncommutativity of observables, it does not reduce to it, and they are different, as I will explain and show by concrete examples in section \sref{s:non_commutativity}. This article is an attempt to reduce most of the known strange features of Quantum Mechanics to a single one. As I will show, this includes complementarity, quantum uncertainty, EPR paradox, Bell's theorem, quantum contextuality and quantum correlations.

\section{The essence of quantumness}
\label{s:QMEssence}

\subsection{The postulates of Quantum Mechanics}
\label{s:QMprinciples}

For the purpose of this article it is enough to rely on a well known formulation of Quantum Mechanics \cite{Dir58,vonNeumann1955foundations}, which I remind briefly.

A {\em quantum system} has associated a {\em Hilbert space} $\hilbert$. The {\em state} of the system is represented by a vector $\ket{\psi}$ in $\hilbert$.
Its {\em time evolution} is governed by a unitary operator $U(t)\in\U(\hilbert)$, by
\begin{equation}
\label{eq:unitary_evolution}
\ket{\psi_t}=U(t)\ket{\psi_0}.
\end{equation}

An {\em observable} is a Hermitian operator $\mc{\hat O}$, and can be written as
\begin{equation}
\label{eq:observable_decomposition}
\mc{\hat O} = \sum_\lambda\lambda\Prj_\lambda,
\end{equation}
where $\Prj_\lambda$ are the projection operators onto the {\em eigenspaces} $\hilbert_\lambda$, indexed by the {\em eigenvalues} $\lambda$.
The Hilbert space $\hilbert$ admits the orthogonal decomposition
\begin{equation}
\label{eq:orthogonal_decomposition}
\hilbert = \bigoplus_{\lambda}\hilbert_\lambda,
\end{equation}
and $\hilbert_\lambda=\Prj_\lambda\hilbert$.
Any {\em observation} or {\em measurement} of the observable $\mc{\hat O}$ of the system whose state is $\ket{\psi}$ is governed by the following postulates, which were distilled from the original Born rule \cite{Dir58,vonNeumann1955foundations}:

\textbf{\PProj}. {\em The projection postulate.}
An observation finds the system in a state obtained by projecting the state before observation, $\ket{\psi}$, on one of the eigenspaces of $\mc{\hat O}$, and the {\em outcome} is the corresponding eigenvalue $\lambda$.

There are also more general measurements, named POVMs (from {\em positive operator valued
measure}), but we will not discuss them here, since they would complicate the exposition without telling something that is not already implicit in {\PProj}.

\textbf{\PProb}. {\em The probability rule.}
The probability that the outcome is the eigenvalue $\lambda$ is
\begin{equation}
\label{eq:born_probability}
p_\lambda = \bra{\psi}\Prj_\lambda\ket{\psi}.
\end{equation}

Although these two postulates are often referred together as ``the Born rule'', I prefer to keep them distinct, because they have distinct roles in the understanding of quantumness.

The {\em expectation value} of $\mc{\hat O}$ is
\begin{equation}
\label{eq:expectation_value_basis}
\expectation{\mc O}_\psi := \sum_\lambda \lambda p_\lambda,
\end{equation}
which can be written as
\begin{equation}
\label{eq:expectation_value}
\expectation{\mc O}_\psi = \bra{\psi}\mc{\hat O}\ket{\psi}.
\end{equation}

\subsection{Entanglement}

If a system consists of two or more subsystems, its Hilbert space $\hilbert$ is the tensor product of the Hilbert spaces $\hilbert_i$ of the subsystems,
\begin{equation}
\label{eq:tensor_product}
	\hilbert=\bigotimes_i\hilbert_i.
\end{equation}

When the Hilbert space can be represented as a tensor product, any vector $\ket{\psi}\in\hilbert$ can be represented as a linear superposition of tensor products of vectors from the spaces $\hilbert_i$. If the state $\ket{\psi}=\bigotimes_i\ket{\psi}_i$, where $\ket{\psi}_i\in\hilbert_i$, it is called {\em separable}, otherwise, it is {\em entangled}. 

This may happen when the system represented on the Hilbert space $\hilbert$ is composed of two or more subsystems, represented on $\hilbert_i$. The most known example is when a system is composed of more particles.

But entanglement can also exist between the path and the spin of a particle. For example, the state of an electron which passes through a Stern-Gerlach device becomes a superposition

$$\frac{1}{\sqrt{2}}\(\ket{\uparrow}\ket{\tn{upper trajectory}} + \ket{\downarrow}\ket{\tn{lower trajectory}}\).$$

In the Stern-Gerlach experiment, by detecting the path, we deduce the spin. Similarly, there is entanglement between path and polarization \cite{MQKJZ2009PathPolarizationEntanglement}. The path-spin entanglement is genuine entanglement and can be converted to entanglement between particles \cite{AMHP2010SwappingPathSpin}.

We see that entanglement has nothing intrinsically related to nonlocality or even to position. Entanglement can be present in quantum systems which don't have as degrees of freedom the position, which is not intrinsically quantum mechanical. Understanding this can clarify part of the usual confusion concerning Quantum Mechanics, as we shall see for example in the section about the EPR paradox.

\subsection{The Tension Principle}
\label{s:TensionPrinciple}

The postulates recalled in \sref{s:QMprinciples} have as consequences the plethora of phenomena which make QM so different from the classical world. Quantum correlations, nonlocality, contextuality, all these points of tension between QM and classical physics, are obviously consequences of the postulates of QM itself.

In the following I want to propose a principle which aims to concentrate the essence of the quantumness in ``one clear, simple sentence''.

According to {\PProj}, an observation {\em constrains} the system to be in an eigenstate of the observable. The evolution of the state vector is governed by the {\schrod} equation, and we can say that the observation determines its initial conditions. But the observation not quite finds the initial conditions, it actually constrains them to be one of its eigenvectors. This is already evident in the {\PProj} principle, but let's reformulate it, for clarity purposes.

\textbf{\PProj'}. The observation {\em constrains} the system to be an eigenstate of the observable.

Different observables impose different constraints on the state of the system, which may be incompatible, in the sense that the same state can't satisfy all of them simultaneously. The aim of this article is to show that this tension between the different constraints is at the root of major typically quantum phenomena.

\textbf{\TENS}. {\em The tension principle.} Quantumness is caused by the tension between the constraints imposed by incompatible observations.

This tension is essential, and follows directly from {\PProj}. The ``tension'', the degree of incompatibility between two observables $\hat A$ and $\hat B$, is given by the commutator
\begin{equation}
\label{eq:commutator}
[\hat A,\hat B]=\hat A\hat B - \hat B \hat A.
\end{equation}

The {\TENS} principle becomes manifest when the settings are such that there is a tension between noncommuting observables. As we shall see in section \sref{s:EPR}, sometimes the observables are apparently commutative, yet the tension is still present. But even in that case, as we will see, noncommutativity is present. As we will explain in section \sref{s:non_commutativity}, noncommutativity doesn't always generate tension, and the two concepts are not identical.

This tension is in fact at the root of the probability rule {\PProb} as well. Assume that we first performed the observation $\hat A$, and found the system in an eigenstate $\ket{\psi}$ of $\hat A$. If the second observation $\hat B$ doesn't commute with $\hat A$, then how can the system be also an eigenstate of $\hat B$? In order to be, it has to be projected. But on which eigenspace of $\hat B$ should it be projected? It will project on any of these eigenspaces, with a given probability. 

The probability rule {\PProb} seems to apply to a single observable, so where is the tension? The tension is between the preparation of the system in an eigenstate of the observable $\hat A$, and the measurement of the observable $\hat B$. While at this point {\PProb} doesn't seem to be different from classical probabilities, we will see that it is responsible for the very nonclassical correlations.

In the following I will argue that whenever we have a deviation of the predictions of QM from those a classical theory would be able in principle to make, the deviation is caused by a tension between the constraints imposed by the observables.

\section{Manifestations of the Tension Principle}
\label{s:manifestations}

\subsection{Complementarity and uncertainty}
\label{s:complementarity_uncertainty}

Probably the first known nonclassical feature of Quantum Mechanics was the point-particle--wave duality. A particle behaves sometimes like a wave, and sometimes like a classical material point, depending on whether we measure the position or the momentum.

Bohr's {\em complementarity principle} states that the wave aspect and the point-particle aspect cannot be observed simultaneously. This corresponds to the fact that the two observables $\hat{x}$
and $\hat{p}_x$ decompose differently the Hilbert space. Hence, Bohr's complementarity is a direct consequence of the tension principle {\TENS}.

Heisenberg's uncertainty principle \cite{Heisenberg1927Uncertainty,Kennard1927Uncertainty,Weyl1928GruppentheorieQuantenmechanik} states that
\begin{equation}
\label{eq:heisenberg_uncertainty}
\sigma_{x,\psi}\sigma_{p_x,\psi}\geq\frac{\hbar}{2},
\end{equation}
where $\sigma_{\mc O,\psi}$ denotes the {\em standard deviation} of the operator $\mc{\hat O}$, defined as
\begin{equation}
\label{eq:standard_deviation}
\sigma_{\mc O,\psi} := \sqrt{\expectation{\mc{\hat O}^2}_\psi-\expectation{\mc{\hat O}}_\psi^2}.
\end{equation}

Robertson \cite{Robertson1929Uncertainty,Schrodinger1930Uncertainty} generalized the uncertainty principle to any two observables $\hat A$ and $\hat B$, and $\ket{\psi}$ in their common domain:
\begin{equation}
\label{eq:robertson_uncertainty}
\sigma_{A,\psi}\sigma_{B,\psi}\geq\frac{1}{2}\abs{\expectation{[\hat A,\hat B]}_\psi}.
\end{equation}

For the observables $\hat{x}$ and $\hat{p}_x$ we recover \eqref{eq:heisenberg_uncertainty}, since
\begin{equation}
[\hat{x},\hat{p}_x]=i\hbar.
\end{equation}

From \eqref{eq:expectation_value_basis} we see that the uncertainty principle follows directly from the tension principle and the probability rule \PProb.

Complementarity and uncertainty were the first hints that quantum mechanics is very different from classical theories. However, they can be emulated by classical theories, provided that we add variables that are not observable (hence called ``hidden variables''). However, in the following we shall see that there are other features, which are not local and depend on the measurements settings. These can be emulated only by hidden variables theories which are nonlocal and depend on the context.

\subsection{EPR and Bell's theorem}
\label{s:EPR}

Consider a system made of two subsystems, $\hilbert=\hilbert_A\otimes\hilbert_B$. Suppose that the state $\ket{\psi}\in\hilbert$ is entangled, and we observe the state of the two systems, where $\mc{\hat O}_A$ and $\mc{\hat O}_B$ are the observables.
This is just a particular case of an observable $\mc{\hat O}$ on the Hilbert space $\hilbert$, where $\mc{\hat O}=\mc{\hat O}_A\otimes\mc{\hat O}_B$, and the Born rule ({\PProj} \& {\PProb}) leads to the expectation value  \eqref{eq:expectation_value} 
\begin{equation}
\label{eq:expectation_value_EPR}
\expectation{\mc O}_\psi = \bra{\psi}\mc{\hat O}_A\otimes\mc{\hat O}_B\ket{\psi}.
\end{equation}
If we interpret the probability distribution determined by {\PProb} in terms of $\hilbert_A$ and $\hilbert_B$, we find that the outcomes of the two observations are correlated, and the correlation is given by the expectation value of the product of the outcomes on the two sides.

The observables $\mc{\hat O}_A$ and $\mc{\hat O}_B$ act on the subsystems $\hilbert_A$ and $\hilbert_B$. They are equivalent to the observables $\mc{\hat O}_A\otimes {\hat I}_{B}$ and ${\hat I}_{A}\otimes\mc{\hat O}_B$, which act on $\hilbert$ and commute. If they commute, then where is the tension present here?

Let's take as an example the EPR experiment \cite{EPR35}, in the version with spins, used by Bohm \cite{Bohm51} and Bell \cite{Bel64}, where we have two qubits entangled in a singlet state
\begin{equation}
\label{eq:bell_singlet}
\ket{\psi}=\frac{1}{\sqrt 2}\(\ket{\uparrow}_A\ket{\downarrow}_B-\ket{\downarrow}_A\ket{\uparrow}_B\),
\end{equation}
and the observables represent the spin along some directions. Suppose the observable $\mc{\hat O}_A$ has the form
\begin{equation*}
\mc{\hat O}_A=\frac{1}{2}\(\ket{\nearrow}_A\bra{\nearrow}_A+\ket{\swarrow}_A\bra{\swarrow}_A\).
\end{equation*}
Its eigenstates are the vectors $\ket{\nearrow}_A$ and $\ket{\swarrow}_A$. Then, the singlet state \eqref{eq:bell_singlet} can be written as
\begin{equation}
\label{eq:bell_singlet_diag}
\ket{\psi}=\frac{1}{\sqrt 2}\(\ket{\nearrow}_A\ket{\swarrow}_B-\ket{\swarrow}_A\ket{\nearrow}_B\).
\end{equation}
Therefore, the observable $\mc{\hat O}_A$ not only singles out the directions $\ket{\nearrow}_A$ and $\ket{\swarrow}_A$ in $\hilbert_A$, but also the directions $\ket{\nearrow}_B$ and $\ket{\swarrow}_B$ in $\hilbert_B$. Let $\mc{\hat O}_B'$ be the observable having as eigenstates these directions,
\begin{equation*}
\mc{\hat O}_B'=\frac{1}{2}\(\ket{\nearrow}_B\bra{\nearrow}_B+\ket{\swarrow}_B\bra{\swarrow}_B\).
\end{equation*}
Then, $[\mc{\hat O}_A\otimes {\hat I}_{B},{\hat I}_{A}\otimes\mc{\hat O}_B']=0$. The observables $\mc{\hat O}_A$ and $\mc{\hat O}_B$ are incompatible if and only if $[\mc{\hat O}_B,\mc{\hat O}_B']\neq 0$, and here is where the tension is present.

In terms of the Choi isomorphism, the state \eqref{eq:bell_singlet}, when expressed in the basis $\(\ket{\nearrow}_A,\ket{\swarrow}_A\)$, determines an isomorphism
\begin{equation}
\label{eq:bell_isomorphism}
\ket{\nearrow}_B\bra{\nearrow}_A+\ket{\swarrow}_B\bra{\swarrow}_A
\end{equation}
between the spaces $\hilbert_A$ and $\hilbert_B$. The observable $\mc{\hat O}_B'$ can be obtained from $\mc{\hat O}_A$ by the isomorphism \eqref{eq:bell_isomorphism}. It is true that this isomorphism depends on the basis, but if we use another basis, we obtain an observable with the same eigenspaces, so this doesn't affect the reasoning. This identification of the space $\hilbert_A$ and $\hilbert_B$ leads, in the case when $[\mc{\hat O}_B,\mc{\hat O}_B']\neq 0$, to a tension between the observables $\mc{\hat O}_A$ and $\mc{\hat O}_B$. If $\mc{\hat O}_B$ and $\mc{\hat O}_B'$ commute, there is no tension, and the correlations are classical.

There is another way to see this tension, as taking place between the preparation and the two observations. To prepare the system in a singlet state, the observable must have as nondegenerate eigenstate the singlet state \eqref{eq:bell_singlet}. Any such observable is incompatible with $\mc{\hat O}_A\otimes\hat I_B$ and $\hat I_A\otimes\mc{\hat O}_B$, which don't admit the singlet state as eigenstate. There is a tension between the preparation in a singlet state and the observables $\mc{\hat O}_A$ and $\mc{\hat O}_B$. Also, there is a tension between $\mc{\hat O}_A$ and $\mc{\hat O}_B$, but only if $[\mc{\hat O}_B,\mc{\hat O}_B']\neq 0$.

Therefore, the tension principle {\TENS} is at the root of the nonclassical behavior manifest in the EPR experiment.

\subsection{Quantum contextuality}

The projection postulate {\PProj}, stating that the system is found to be an eigenstate of the observable $\mc{\hat O}$, has a strange feature. If a system is in a definite state before the measurement, it seems to depend on what observable will be measured in the future.
While Bohr resolved this problem by suggesting that there is no reality prior to the observation, this was not compelling for everyone. The reason is that one can easily conceive that there is a pre-assignment of outcomes for each possible observation.

The {\ks} theorem \cite{KochenSpecker1967HiddenVariables} shows that the setup can be chosen so that no pre-assignment of outcomes for each possible observation is possible. Or in fact it depends not only on that observation, but also on any other observations performed together with it.

Assume that a particle contains the information to determine the outcome of any possible observation $\mc{\hat O}$ we can perform on it. Then, the theorem shows, this information should also depend on the {\em context}, \ie on what other observable $\mc{\hat O'}$ we measure together with $\mc{\hat O}$, even if they commute.
Consequently, if one tries to build a hidden-variables theory in which the particle contains the information needed to determine the outcome of any possible observation, one should actually make sure this depends on the other observations too.

The {\ks} theorem is obtained by gathering enough observables, so that the tension between them prevents the possibility to pre-assign outcomes for each possible observation.

The original {\ks} theorem doesn't involve the probability rule {\PProb}, but there are variants which rely on correlations \cite{KCBS2008HiddenVariables}.
Conversely, there are also Bell-type results which follow solely from {\PProj} and not from {\PProb}, for example involving the GHZ state \cite{GHZ1989BeyondBell,GHSZ1990BellWithoutInequalities,Mermin1993HV2Bell}.

\subsection{Quantum correlations}

The correlation between two observables $\mc{\hat A}$ and $\mc{\hat B}$ is defined as the expectation value of their product, 
\begin{equation}
\label{eq:correlation}
C(\hat A,\hat B)_\psi=\expectation{\hat A\hat B}_{\psi}.
\end{equation}

It is interesting how expectation values work in QM. On the one hand, as seen from the equation \eqref{eq:standard_deviation}, they lead to the uncertainty principle \eqref{eq:heisenberg_uncertainty}, and were regarded as proving the limitations of QM, as compared to the potentially infinite precision of measurement in classical mechanics. However, by carefully combining observables, the probability rule {\PProb} can lead to correlations that can't be obtained by classical theories, unless we allow them to violate local realism and independence of context.

The greater the tension between the constraints imposed by the observables, the greater and more nonclassical is the resulting correlation.

Classical correlations are calculated assuming that there are some definite values for the variables. The resulting correlations satisfy some inequalities, which may be violated by the corresponding quantum correlations, because the tension principle doesn't allow the existence of definite values for all the observables. This provides ways to test the predictions of QM as compared to those of classical theories.

If the quantum observations are supposed to be made simultaneously on the same system in the same place, one obtains {\ks}-type inequalities.

If the observations are sequential, one obtains temporal inequalities, or {\leggarg}-type inequalities \cite{LeggettGarg1985QMvsMacrorealism}.
There are some theories which aim to prove that the classical world emerges from the quantum one. For instance, the {\em objective collapse} theories \cite{GRW86}, and the {\em decoherence program} \cite{zurek1981pointer,Zur03a}. But is the classical theory which is supposed to emerge at the macroscopic level realist? The {\leggarg} theorem ruled out macroscopic realism, similar to how the Bell \cite{Bel64} and {\ks} \cite{Bel66,KochenSpecker1967HiddenVariables} theorems ruled out local and noncontextual realism. The {\leggarg} inequality is violated at all quantum levels, and became a prototype for the temporal inequalities.

\subsection{A unified view on correlations}
\label{s:unified_correlations}

Quantum correlations may appear very different: the {\ks}-type correlations refer to the context, the {\leggarg}-type correlations are temporal, and the Bell-type correlations are spatial. However, they all have something in common: the classical inequalities are obtained by similar algebraic calculations, and all the quantum correlations follow from the tension between the constraints imposed by the noncommuting observables. This suggests that they may be more strongly related.

Suppose we start with a set of observables for the same system, so that there is a tension between the constraints they impose. For some states the quantum correlations violate the inequalities which are obeyed by the corresponding classical correlations. We obtained a {\ks}-type scenario. Now, let's assume the same observations are performed in a temporal order. For example, the KCBS inequality \cite{KCBS2008HiddenVariables} can be translated in a temporal scenario easily. We obtain a {\leggarg}-type scenario. Hence, the temporal, or {\leggarg}-type scenarios are also of the {\ks}-type,
\begin{equation}
\label{eq:inequalities_lg_ks}
\tn{\leggarg}\subset\tn{\ks}.
\end{equation}

Assume now that we are in a temporal scenario, and the observed system is composite, and the observables act on one of the subsystems or the other. The observables corresponding to different subsystems commute, so it seems there is no tension. However, as explained in \sref{s:EPR}, the state in which the composite system is prepared forces relations between the Hilbert spaces of the subsystems. This means that what we thought to be compatible observables, for entangled systems are in fact incompatible. Alternatively, the tension can be seen as taking place between the preparation of the entangled system and the other observables.
In the Bell-type scenario, space doesn't matter, because is not part of the postulates which led to the Bell correlations. What matters is that a tensor product and entanglement are involved, otherwise there is no difference from a general {\leggarg}-type scenario. Therefore,
\begin{equation}
\label{eq:inequalities_bell_lg}
\tn{Bell}\subset\tn{\leggarg}.
\end{equation}

Bell-type inequalities are just temporal inequalities involving commuting observables, applied to entangled states. They appear different from the temporal inequalities because the observables $\mc{\hat O}_A$ and $\mc{\hat O}_B$ appear to act on different particles, which can be separated in space. But the presence of the space separation is only intended to emphasize the tension, by preventing local interactions to account for the correlations between the outcomes of the observables $\mc{\hat O}_A$ and $\mc{\hat O}_B$.

While the general temporal correlations are bound easier by using {\em semidefinite programming} \cite{Guhne2013BoundTimeQCorrelations}, this is more difficult for the Bell-type inequalities, because one has to restrict the methods to observables that commute \cite{Navascues2007BoundingQCorrelations,Navascues2008SemidefiniteQCorrelations}.

We can now summarize the relations \eqref{eq:inequalities_lg_ks} and \eqref{eq:inequalities_bell_lg} between the three types of inequalities:
\begin{equation}
\label{eq:all_inequalities}
\tn{Bell}\subset\tn{\leggarg}\subset\tn{\ks}.
\end{equation}

However, this hierarchy doesn't prohibit the existence of procedures to convert {\ks} theorems into Bell's theorems \cite{heywood1983nonlocality,Mermin1990UnifiedNHV,Mermin1990ExtremeEntanglement,Mermin1993HV2Bell,Cabello2001AllVsNothing}.

Quantum correlations are stronger than classical correlations. A classical mechanism would allow this amount of correlations work only if the two subsystems would interact or exchange information. Classical multivariate probability distributions are {\em joint probability distributions}. But in QM, it is not always possible to assign definite values to all variables simultaneously. Therefore, as shown by A. Fine \cite{Fine1982JointDistributions,Fine1982HVJointBell}, for Bell-type scenarios there is no joint probability distribution which gives the same correlations as QM. Any joint probability distribution should obey Bell's inequality, while QM violates it. 
The idea that quantum correlations are characterized by the nonexistence of a joint probability distribution applies also to {\ks} type correlations \cite{MalleyFine2005NoncommutingObservablesLocalRealism,KCBS2008HiddenVariables}, and temporal correlations \cite{Fritz2010CHSHCorrelations,UnifiedCorrelations2013}.
Because quantum correlations are just expectation values of products of observables, the joint probability distributions exist when the involved observables commute. This suggests that the source of quantumness is the noncommutativity of observables \cite{Malley2004CommuteSimultaneously,MalleyFine2005NoncommutingObservablesLocalRealism}. However, as we shall see, while the tension requires noncommutativity, noncommutativity doesn't always lead to tension.

We have seen thus that the tension is also at the root of quantum probabilities.

\section{Tension and noncommutativity}
\label{s:non_commutativity}

From the dawn of Quantum Mechanics, the fact that observables may not commute has been seen as the key feature of the phenomena characteristic to quantum world. And this is partly true, because noncommutativity can be found at the root of all these phenomena. In the meantime, new strange quantum phenomena were discovered, which seemed to take place also when noncommutativity is absent, like the EPR paradox. Although in general the EPR paradox is presented as not involving noncommutativity, because the observables $\mc{\hat O}_A$ and $\mc{\hat O}_B$ commute, we have seen in \sref{s:EPR}, that here noncommutativity is present too. Similarly, contextuality involves compatible observations, however, even in the original {\ks} theorem \cite{KochenSpecker1967HiddenVariables} are considered noncommuting sets of observables that commute. Hence, noncommutativity is a signature of quantumness.

So, why trying to explain quantumness in terms of the tension principle, and not simply in terms of noncommutativity? Here are some differences between the two.

First, there are cases when even commuting observables give rise to tension. Consider for instance that we make an observation, then the system evolved in time according to the {\schrod} equation, and then we repeat the observation. During the time evolution, the system may have changed and is no longer in an eigenstate of the observable, so the tension appears. Of course, this can be easily resolved by switching to the Heisenberg picture. In the Heisenberg picture the two observables no longer commute. Commutativity of observables differs in the {\schrod} and Heisenberg pictures, and it is the latter that is more relevant to this.

More important, noncommutativity doesn't necessarily generate tension. Consider the case when two incompatible observations are made, but they share common eigenvectors. If the first observation projects the observed system into one of these common eigenvectors, then the second observation leaves it unchanged. So here we have noncommutativity, but not tension, and the sequence of observations behaves similarly to the classical ones. 

We see that noncommutativity is relevant because it leads to the tension, but it simply doesn't matter when it doesn't lead to tension. The tension principle captures better quantumness, and therefore has a better explanatory power than noncommutativity.

\section{Quantum Mechanics in relation to spacetime}
\label{s:spacetime}

We have seen so far that the tension principle is responsible for the internal weirdness of QM. Yet, when we take into account that the particles have to live in a spacetime, this weirdness can be amplified to become more evident. Then, by comparing quantum phenomena with nonquantum ones taking place in spacetime, QM seems even stranger.

An example of such amplification is present in the EPR experiment. 
We have seen in sections \sref{s:EPR} and \sref{s:unified_correlations} that EPR experiment can be easily understood from the principles of QM, if we forget about space-like separation of the two entangled particles. This space-like separation is what makes QM appear incomplete, like Einstein, Podolsky and Rosen concluded in their paper \cite{EPR35}.

The idea of modifying an experiment in order to emphasize or amplify the weirdness of Quantum Mechanics was also used by J.A. Wheeler in his famous ``delayed choice experiment'' \cite{Whe78}. In this experiment, the observer chooses whether to measure ``which path'' of ``both ways'' after the photon passed through the double-slit. Depending on this choice, the photon follows either a path or the other, respectively both paths, as if it seems to anticipate what the observer will choose to measure. This delay of the choice made by the observer emphasizes the fact that there is no way the experimental setup influenced through local interactions the interaction of the photons with the double-slit. To make the things worse, in \cite{ionicioiu2011quantum_delayed_choice} is proposed an experiment in which the choice is postponed even after the measurement took place.

The internal weirdness of QM can be, if not completely understood, at least acceptable. However, the external weirdness, that which is manifest when we try to make sense in the same time of both QM and Relativity seems more acute. It is at this edge where concepts like {\em nonlocality}, {\em realism}, faster than light {\em signaling}, {\em ontic} \vs {\em epistemic} {\etc}, become important. While I avoided to discuss them in order to simplify the exposition and make it clearer that the tension is the core of quantumness, this omission does not sweep under the carpet any of the internal problems of QM. But regarding the external ones, we would want to make sense of QM and the macroscopic world, which appears to be classical. We also want to be able to think at QM and Special or, if possible, General Relativity as describing the same world, not completely different universes, with different, incompatible sets of laws. This is why various attempts to interpret, reformulate, reconstruct or modify QM are so important.

To do this, it may be useful to identify the essence of quantumness, as a central point which allows us to solve simultaneously as many conflicts as possible. In this paper I brought arguments that this essence is the tension principle. This makes quantum phenomena so different than the other phenomena. So, in order to resolve the problems which appear at the interface between QM and other fields, we have to resolve this tension.

This is why hidden-variable theories tried to modify or complete QM, to make it compatible with the notions of causality inherited from Classical Mechanics, which are more intuitive and easier to accept. Actually, we can say that most interpretations try to resolve in one way or another the internal tensions of QM, as well as the tension between QM and the classical or special-relativistic notion of causality.

I will mention a particular proposal which attempts to solve precisely the tension identified in this article, without modifying QM, and even without breaking the unitary evolution governed by the {\schrod} equation \cite{Sto08b,Sto08f,Sto12QMa}. The first idea on which this solution is based is that if there is only one observable involved, or if all observables commute, then there is no tension between the constraints on the observed system. Even in this case, in order for the explanation to work without breaking the unitary evolution, the initial conditions of the observed system have to depend on those of the measurement apparatus \cite{Sto12QMb}. But when incompatible observables are present, and they impose constraints which are in tension, this is not enough.

Here comes the second idea, which is that whenever a system is observed, it becomes entangled with the measurement apparatus. Hence, the next measurement acts on a larger Hilbert space. This in its turn enlarges the Hilbert space by entangling the system with the corresponding apparatus. Each new measurement or interaction of the system adds new degrees of freedom, and enlarges the observed Hilbert space. But in a larger Hilbert space, there is more room for the observables to commute. And if they do commute, the tension is resolved. Of course, it is an open problem that the injection of degrees of freedom by each new measurement is enough to make the observables commute, but this argument shows that it is possible, at least in principle, to resolve the tension. This solution is compatible with the causality from both Special and General Relativity, and in fact the block universe view of Relativity seems to be a better frame to understand how there can be solutions of the {\schrod} equation that satisfy the constraints imposed by incompatible observations \cite{Sto12QMc,Sto13b}.

\section{Conclusion}

We have seen that the many features of Quantum Mechanics that appear counterintuitive and paradoxical, can be seen as various manifestations of a single principle -- the tension principle. This principle states that they are caused by the tension between the constraints imposed by incompatible observations. The tension principle itself is counterintuitive.

The tension principle is not introduced as a completion of the other principles of Quantum Mechanics, nor it is not used to derive them, being in fact implicit in them. The role of emphasizing the tension principle is to show that, if we have it in mind when thinking at the paradoxes of Quantum Mechanics, it contains their essence. We can say that the quantum paradoxes have their root in one single paradox, namely that the constraints appear to be contradictory. But maybe Bohr's saying applies here too: {\em Contraria non contradictoria sed complementa sunt}. Even if this principle gathers the essence of other typically quantum phenomena, it would be useful to understand how the tension is resolved. This remains an open problem, and can be seen as the starting point of possible interpretations and reformulations of Quantum Mechanics.

{\bf Acknowledgments}

The author cordially thanks 
Radu Ionicioiu and Florin Moldovea\-nu
for very helpful comments and suggestions.


\end{document}